\documentclass[notitlepage,superscriptaddress,showpacs,nobalancelastpage,twocolumn,aps,prl,floatfix,longbibliography]{revtex4-2}

\usepackage[english]{babel}
\RequirePackage[T1]{fontenc}
\RequirePackage{times}
\usepackage{siunitx}
\usepackage[italicdiff]{physics}
\usepackage{amsfonts}
\usepackage{amsmath}
\usepackage{amssymb}
\usepackage{graphicx}
\usepackage{xspace}
\usepackage{array}
\usepackage[hidelinks]{hyperref}
\usepackage{xcolor}
\usepackage{bm}
\usepackage{verbatim}
\usepackage{tikz}
\usepackage{float}
\allowdisplaybreaks[3]

\makeatletter
\AtBeginDocument{\let\LS@rot\@undefined}
\makeatother

\usetikzlibrary{calc,3d}
\usetikzlibrary{arrows}
\usetikzlibrary{decorations.pathmorphing}

\usepackage[normalem]{ulem}

\begin{document}

\title{Acoustic Tweezers for Magnetic Skyrmions}

\author{Chongzhou Wang}
\affiliation{State Key Laboratory of Surface Physics and Institute for Nanoelectronic Devices and Quantum Computing, Fudan University, Shanghai 200433, China}
\affiliation{Department of Physics, Fudan University, Shanghai 200433, China}

\author{Weichao Yu}
\email{wcyu@fudan.edu.cn}
\affiliation{State Key Laboratory of Surface Physics and Institute for Nanoelectronic Devices and Quantum Computing, Fudan University, Shanghai 200433, China}
\affiliation{Zhangjiang Fudan International Innovation Center, Fudan University, Shanghai 201210, China}

\date{\today}

\begin{abstract}
Current methods for driving magnetic skyrmions predominantly translate ensembles as a whole, lacking single-particle selectivity. Here, we propose an ``acoustic tweezer'' that deterministically traps and routes individual skyrmions using spatially extended acoustic beams. We reveal that spatially confined longitudinal waves carry nontrivial phonon spin, inducing a magnetoelastic field whose chirality is locked to the acoustic spin texture. This generates polarity-selective radiation forces, distinct from conservative gradient forces, that attract skyrmions to local phonon spin maxima. Intersecting orthogonal beams create reconfigurable attractive points for adiabatic, deterministic manipulation. Our global-field-local-interaction paradigm establishes a non-destructive, on-chip route for high-precision topological spintronics.
\end{abstract}

\maketitle

\textit{Introduction}--- Precise manipulation of microscopic entities constitutes a central objective spanning fundamental physics, biomedicine, and advanced materials science. Optical and acoustic tweezers have emerged as mature paradigms to address this challenge \cite{Grier2003,Gao2017,Ozcelik2018}. Optical tweezers leverage the mechanical, thermal, and electromagnetic effects of optical fields to achieve spatiotemporally resolved manipulation across diverse targets, including living cells \cite{Čižmár2010,Memmolo:15}, metallic nanoparticles \cite{Lin2018}, and dielectric nanostructures \cite{PhysRevLett.79.645,Krishnan2010}. Acoustic tweezers harness phased transducer arrays \cite{Orazbayev2024,Marzo2015}, standing-wave patterning \cite{doi:10.1073/pnas.1209288109,doi:10.1073/pnas.1504484112,B910595F,doi:10.1073/pnas.1524813113,doi:10.1073/pnas.1813047115,Takatori2016}, and acoustic vortices \cite{Ahmed2016} to facilitate reconfigurable mechanical field geometries with lower input power density. Beyond the manipulation of discrete physical particles, the control of topologically stabilized magnetic textures, including skyrmions, domain walls, and vortices, has emerged as a distinct research frontier. The topological protection of these textures enables robust information carriers whose deterministic manipulation would unlock new functionalities for spintronic architectures.

Despite the proven efficacy of magnetic fields \cite{Beach2005,PhysRevLett.96.197207,Moon2016,Wang_2017}, spin-polarized currents \cite{PhysRevLett.92.086601,Birch2024}, and spin waves \cite{PhysRevB.107.224418,10.1063/5.0284205} in transferring spin torque to drive magnetic textures, achieving spatially and temporally resolved capture with reconfigurable manipulation capabilities remains experimentally elusive. Recent methodologies, including localized laser-induced thermal gradients \cite{PhysRevApplied.19.044036,Kim2025,Mochizuki2014}, orbital angular momentum transfer from spin waves \cite{PhysRevLett.124.217204}, and surface acoustic wave excitations \cite{Rivelles2025,w75m-hlsr,PhysRevB.107.144421,10.1063/5.0207929,Miyazaki2023,Chen2023,Yang2024}, have demonstrated controlled skyrmion motion over extended distances. However, these approaches predominantly drive multiple skyrmions collectively, and deterministic addressability of individual topological defects along arbitrary trajectories or at specific local coordinates has not yet been realized.

To bridge this gap, we exploit the local topological properties of acoustic fields, specifically the phonon spin angular momentum. Recent studies have established that phonon spin governs the intricate polarization states of acoustic fields \cite{doi:10.1073/pnas.1808534115,PhysRevLett.131.136102,Yuan2021}, and the magnetoelastic coupling strength exhibits a pronounced dependence on phonon spin orientation \cite{doi:10.1126/sciadv.ado2504,doi:10.1126/sciadv.abb1724,Wang2026}. We propose an ``acoustic tweezer'' for magnetic skyrmions based on this localized interaction. Although the acoustic wave propagates macroscopically through the magnetic medium, the magnetoelastic coupling is highly localized and chirality-dependent. Magnetic skyrmions, classified into binary polarity states based on their core magnetization, engage in polarity-selective attraction or repulsion with the phonon spin texture. This localized potential landscape enables the precise manipulation of individual skyrmions, overcoming the limitations of global driving fields. Our approach inverts the conventional logic of localized-field-to-localized-target manipulation: we employ spatially extended fields whose internal topological structure inherently generates localized interaction zones. This global-field-local-interaction scheme provides a viable route to deterministic single-skyrmion control within dense ensembles \cite{Zhao2024}.

\textit{Phonon spin and magnetoelastic chirality}--- An ideal plane longitudinal wave carries zero phonon spin and exhibits strictly linear polarization throughout space. However, for spatially confined Gaussian longitudinal acoustic beams, the transverse amplitude decay inhomogeneity qualitatively alters the polarization state \cite{doi:10.1073/pnas.1808534115}. In an isotropic elastic medium, any displacement field must satisfy the irrotational condition $\nabla \times \mathbf{u} = 0$ for pure longitudinal modes, which in the two-dimensional $x$-$y$ plane requires the constraint $\partial u_y/\partial x = \partial u_x/\partial y$. When the longitudinal displacement takes the form of a plane wave with a transverse Gaussian envelope, $u_x(x,y,t) = u_0 e^{-y^2/\delta^2} \cos(kx - \omega t)$, where $u_0$ denotes the peak displacement amplitude, $\delta$ represents the beam waist, $k$ is the acoustic wavenumber, and $\omega$ is the angular frequency, the irrotational constraint strictly induces a transverse displacement component $u_y(x,y,t) = -u_0\frac{2y}{k\delta^2}e^{-y^2/\delta^2}\sin(kx-\omega t)$. The spatial integration over $x$ inherently introduces a $\pi/2$ phase shift between the longitudinal and transverse components, while the $y$-derivative extracts an odd spatial parity factor. This mathematical structure guarantees that the local lattice vibration is elliptically polarized, with the handedness reversing across the beam axis.

\begin{figure}
  \includegraphics[width=0.45\textwidth]{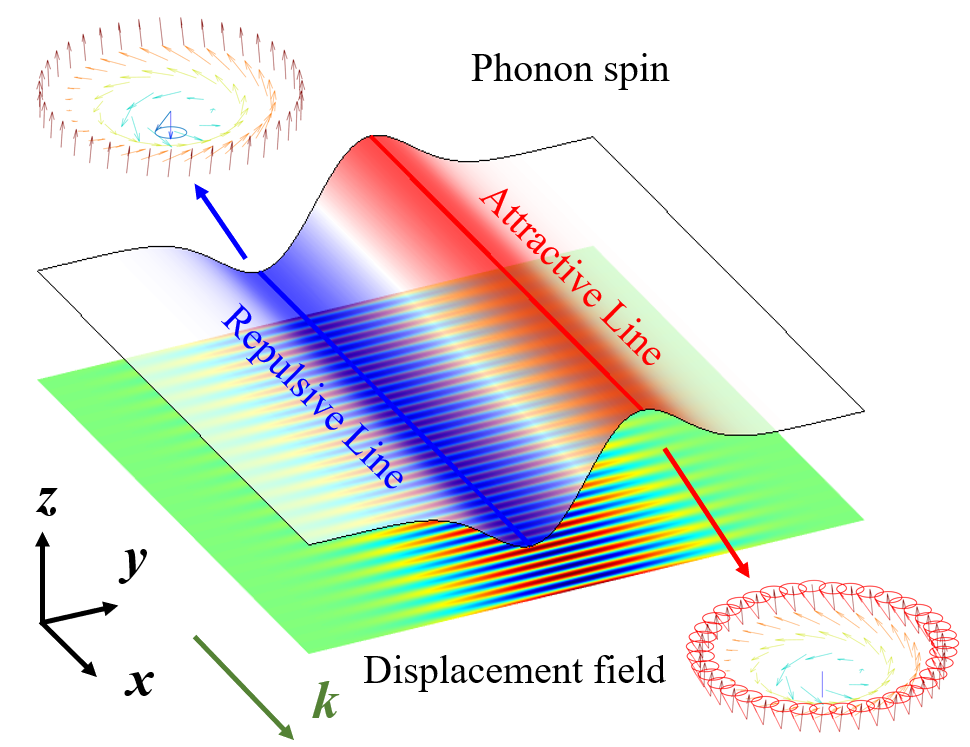}
  \centering
  \caption{\label{fig_skyrmion:fig1} Schematic of phonon-spin-mediated skyrmion interaction. A longitudinally polarized Gaussian acoustic beam propagates along $+\hat{\mathbf{x}}$ (light blue arrow). Transverse amplitude decay induces localized phonon spin angular momentum along $\hat{\mathbf{z}}$ (gradient-colored iso-surfaces), forming an attractive line (red, positive phonon spin) on the upper side and a repulsive line (blue, negative phonon spin) on the lower side. A Bloch skyrmion ($Q = -1$) exhibits polarity-selective responses: it is attracted toward the attractive line and repelled from the repulsive line, dictated by the signed coupling between skyrmion polarity and phonon spin orientation.}
\end{figure}

The resulting phonon spin angular momentum density $\mathbf{S}_{\text{p}} = \rho\langle\mathbf{u}\times\dot{\mathbf{u}}\rangle$ \cite{1670231018604-1340368688,doi:10.1073/pnas.2411427121}, where $\rho$ is the mass density, characterizes the local elliptical polarization state of lattice vibrations and is distinct from the phonon orbital angular momentum associated with wavefront topology. For the Gaussian beam, this evaluates to:
\begin{equation}
    \mathbf{S}_{\text{p}} = \frac{2\rho u_0^2 c_{\text{l}} y}{\delta^2}\exp\left(-\frac{2y^2}{\delta^2}\right)\hat{\mathbf{z}},
\end{equation}
where $c_{\text{l}} = \omega/k$ is the longitudinal sound velocity. The phonon spin density is an odd function of $y$, demonstrating that counter-oriented phonon spin domains emerge naturally on opposite sides of the propagation axis, as illustrated in Fig.~\ref{fig_skyrmion:fig1}.

Spin-lattice interactions mediate magnetoelastic coupling between the elastic and magnetic subsystems. For a cubic lattice, the magnetoelastic energy density is $\mathcal{E}_{\text{me}} = b_1 \sum_i \varepsilon_{ii} m_i^2 + b_2 \sum_{i \neq j} \varepsilon_{ij} m_i m_j$ \cite{PhysRev.110.836,PhysRevB.86.134415,PhysRevB.102.134417}, where $\varepsilon_{ij} = \frac{1}{2}(\partial_i u_j + \partial_j u_i)$ is the strain tensor, $\mathbf{m}$ is the unit magnetization vector, and $b_1$, $b_2$ are the magnetoelastic coupling coefficients. The dominant in-plane strain components generated by the Gaussian beam are:
\begin{align}
    \varepsilon_{xx} &= -k u_0 e^{-y^2/\delta^2}\sin(kx-\omega t), \notag \\
    \varepsilon_{xy} &= -u_0\frac{2y}{\delta^2}e^{-y^2/\delta^2}\cos(kx-\omega t).
\end{align}
The magnetoelastic effective magnetic field $\mathbf{H}_{\text{mec}} = -\frac{1}{\mu_0 M_s}\frac{\partial \mathcal{E}_{\text{me}}}{\partial \mathbf{m}}$, where $\mu_0$ is the vacuum permeability and $M_s$ is the saturation magnetization, then satisfies:
\begin{align}
    H_{x}^{\text{mec}} &\propto m_x\sin(kx-\omega t) + \text{Sgn}(y)m_y\cos(kx-\omega t), \notag \\
    H_{y}^{\text{mec}} &\propto \text{Sgn}(y)m_x\cos(kx-\omega t) - m_y\sin(kx-\omega t).
\end{align}
The transverse field component $H_{y}^{\text{mec}}$ lags by $\pi/2$ for $y>0$ (right-handed polarization) whereas it leads by $\pi/2$ for $y<0$ (left-handed polarization). To quantify this relationship rigorously, we define the effective field chirality density as $\mathcal{C}_H = \langle H_{x}^{\text{mec}}\dot{H}_{y}^{\text{mec}} - H_{y}^{\text{mec}}\dot{H}_{x}^{\text{mec}}\rangle$. Direct calculation yields:
\begin{equation}
    \mathcal{C}_H \propto \frac{y}{\delta^2}\,e^{-2y^2/\delta^2}\left[k\,m_x^2 - \frac{2}{k\delta^2}\!\left(1-\frac{2y^2}{\delta^2}\right)m_y^2\right],
    \label{eq:CH}
\end{equation}
with complete expression given in Appendix~\ref{sec:chirality}. Crucially, this expression shares the identical spatial envelope $y\,e^{-2y^2/\delta^2}$ and the identical odd-parity factor $\text{Sgn}(y)$ with the phonon spin density $S_{\text{p},z}$ in Eq.~(1), establishing a strict parity correspondence between the effective field chirality and the acoustic phonon spin.

In ferromagnetic systems, magnetic moments undergo exclusively right-handed precession around effective fields. The Gaussian beam generates an effective field with opposite handedness across the beam axis, selectively exciting magnetic moments of differing orientations. For a Bloch skyrmion with core polarity $Q = -1$ (central moment along $-\hat{\mathbf{z}}$, majority magnetization along $+\hat{\mathbf{z}}$), the spatially varying polarization state results in an asymmetric Rayleigh dissipation landscape. The system evolves toward the state of maximum power absorption, which constitutes a stable non-equilibrium attractor in the dissipative phase space. This attractor generates a net time-averaged radiation force, termed by analogy with the scattering (radiation) force in optical tweezers, though here arising from magnetoelastic coupling rather than direct phonon momentum transfer, that attracts the skyrmion toward the attractive line (positive phonon spin polarization) and repels it from the repulsive line (negative polarization domain), as schematically illustrated in Fig.~\ref{fig_skyrmion:fig1}. Reversing the core polarity ($Q = +1$) inverts the selectivity, as dictated by the sign of $Q$.

\begin{figure*}
  \includegraphics[width=0.9\textwidth]{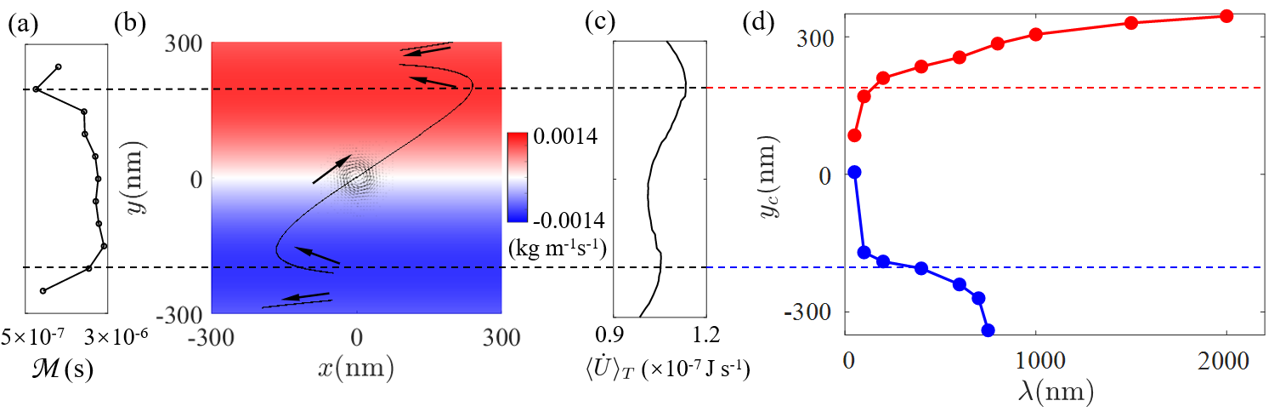}
  \centering
  \caption{\label{fig_skyrmion:fig2} Skyrmion dynamics under a single Gaussian longitudinal acoustic beam propagating along $+\hat{\mathbf{x}}$. (a) Position-dependent effective mass $\mathcal{M}$, reaching a minimum near the attractive line (red) where skyrmion deformation is minimal, and increasing near the repulsive line (blue) and spin extrema due to maximal deformation. (b) Skyrmion trajectory in the phonon spin landscape (color map) showing quasi-static migration toward the attractive line (positive phonon spin region on the upper $y$-axis, red) and away from the repulsive line (negative phonon spin region on the lower $y$-axis, blue). (c) Rayleigh dissipation rate distribution exhibiting extrema that coincide precisely with the phonon spin maxima, confirming polarity-selective magnetoelastic excitation. (d) Wavelength dependence of the attractive line (red dots) and repulsive line (blue dots) positions, demonstrating that the trapping geometry is tunable through acoustic frequency selection.}
\end{figure*}

\textit{Numerical modeling of skyrmion dynamics}--- To validate the analytical predictions, spin dynamics are modeled via the Landau-Lifshitz-Gilbert (LLG) equation \cite{PhysRev.73.155,PhysRevX.5.041049}:
\begin{equation}
    \frac{\partial \mathbf{m}}{\partial t} = -\gamma \mathbf{m} \times \mathbf{H}_{\text{eff}} + \alpha \mathbf{m} \times \frac{\partial \mathbf{m}}{\partial t},
\end{equation}
where $\gamma$ is the gyromagnetic ratio, $\alpha$ is the Gilbert damping parameter, and the total effective field $\mathbf{H}_{\text{eff}} = A_{\text{ex}}\nabla^2\mathbf{m} + K(\mathbf{m}\cdot\hat{\mathbf{z}})\hat{\mathbf{z}} + D(\nabla\times\mathbf{m}) + \mathbf{H}_{\text{mec}}$ incorporates the Heisenberg exchange ($A_{\text{ex}}$), perpendicular uniaxial anisotropy ($K$), and bulk Dzyaloshinskii-Moriya interaction ($D$), together with the magnetoelastic coupling field $\mathbf{H}_{\text{mec}}$ defined above. Without loss of generality, we adopt \cite{PhysRevX.5.041049} $\gamma = 2.21\times10^5\,\text{m/(A}\cdot\text{s)}$, $M_s = 0.194\times10^6\,\text{A/m}$, $A_{\text{ex}} = 0.328\times10^{-10}\,\text{A}\cdot\text{m}$, $K = 0.388\times10^5\,\text{A/m}$, $D = 1.0\times10^{-3}\,\text{A}$, and $\alpha = 0.8$, yielding a skyrmion radius $R_{\text{sk}} \approx 27\,\text{nm}$. The Gaussian acoustic beam is characterized by $u_0 = 5\,\text{nm}$, $\lambda = 600\,\text{nm}$, $\delta = 400\,\text{nm}$, $b_1 = 3.48\times10^5\,\text{J/m}^3$, $b_2 = 6.96\times10^5\,\text{J/m}^3$, and $c_{\text{l}} = 7209\,\text{m/s}$ \cite{PhysRevB.104.014403}. All micromagnetic simulations are performed using the Micromagnetics Module of COMSOL Multiphysics \cite{comsol,Zhang2023,Wang2026}. Since the skyrmion drift velocity is many orders of magnitude below the sound velocity, a one-way coupling scheme is adopted without back-action on the elastic subsystem.

Under Gaussian longitudinal acoustic beam excitation, the skyrmion exhibits a distinctive migration behavior governed by the spatial distribution of phonon spin. As shown in Fig.~\ref{fig_skyrmion:fig2}(b), when a skyrmion is initialized at an arbitrary position in the acoustic field, it undergoes quasi-static motion (i.e., the skyrmion velocity satisfies $|\mathbf{v}| \ll c_{\text{l}}$ and the displacement per acoustic period is negligible compared to $R_{\text{sk}}$) driven by the magnetoelastic force. The phonon spin distribution creates two distinct special regions: an attractive line (red in Fig.~\ref{fig_skyrmion:fig1}) located at a specific position on the positive $y$-axis where the phonon spin reaches its positive maximum, and a repulsive line (blue) at the corresponding negative $y$-axis position with negative maximum phonon spin.

Unlike conservative systems governed by potential energy minimization, our system is a periodically driven dissipative system, where steady-state positions are dictated by the extrema of the Rayleigh dissipation function \cite{PhysRevB.81.014412}. For a skyrmion with downward core magnetization ($Q = -1$), migration into the positive phonon spin region maximizes the phase-locked excitation amplitude of its magnetic moments. This state corresponds to a maximum energy dissipation through Gilbert damping, acting as a stable attractor in the non-equilibrium phase space. Conversely, positioning in the negative phonon spin region leads to polarization mismatch, suppressing excitation and minimizing dissipation, which acts as an unstable repeller. This driven-dissipative mechanism is quantitatively verified by the Rayleigh dissipation formalism \cite{PhysRevB.81.014412}:
\begin{equation}
    \langle\dot{U}\rangle_T = \frac{\alpha\mu_0 M_s d}{\gamma T} \int_0^{T}\mathrm{d}t\int\mathrm{d}^2r\left(\frac{\mathrm{d}\mathbf{m}}{\mathrm{d}t}\right)^2,
    \label{eq:rayleigh}
\end{equation}
where $d$ is the film thickness and $T$ is the acoustic oscillation period. The spatial distribution of the time-averaged dissipation rate [Fig.~\ref{fig_skyrmion:fig2}(c)] exhibits two extrema that coincide precisely with the phonon spin extrema, confirming polarity-selective magnetoelastic excitation.

The skyrmion trajectory exhibits pronounced nonlinearity, contrasting sharply with conventional rigid-body Thiele equation predictions \cite{PhysRevLett.30.230}. This deviation originates from significant acoustic-induced skyrmion deformation under the relatively large displacement amplitude employed to accelerate numerical convergence. The deformation modifies the effective mass of the skyrmion, necessitating a mass-modified generalized Thiele equation. Following the standard derivation from the LLG equation (see Appendix~\ref{sec:thiele} for details), the total force density acting on a moving magnetic texture can be decomposed into gyroscopic, damping, and external contributions. Integrating over the skyrmion volume yields the generalized equation of motion \cite{PhysRevLett.30.230,Mochizuki2014}:
\begin{equation}
    \mathbf{G}\times\mathbf{v} + \alpha\overset{\leftrightarrow}{\mathbf{D}}\cdot\mathbf{v} + \mathcal{M}\dot{\mathbf{v}} = \mathbf{F}_{\text{mec}},
    \label{Thiele Equation}
\end{equation}
where $\mathbf{G} = 4\pi Q\hat{\mathbf{z}}$ is the gyrotropic vector with $Q$ being the topological charge, $\overset{\leftrightarrow}{\mathbf{D}}$ is the damping tensor defined by $D_{ij} = \int\mathrm{d}^2r\,\partial_i\mathbf{m}\cdot\partial_j\mathbf{m}$, $\mathcal{M}$ is the effective skyrmion mass tensor \cite{Mochizuki2014}, and $\mathbf{F}_{\text{mec}}$ is the magnetoelastic driving force acting on the skyrmion centre-of-mass at position $\mathbf{R}$. The inertial term $\mathcal{M}\dot{\mathbf{v}}$ accounts for the finite size and deformation of the skyrmion, introducing initial-velocity-dependent dynamics analogous to Newtonian particles.

Macroscopically, the magnetoelastic driving force $\mathbf{F}_{\text{mec}}$ in Eq.~(\ref{Thiele Equation}) is defined as the volume integral of the force density $\mathbf{f} = -\gamma \sum_j (\nabla m_j) H_{\text{mec}, j}$ over the skyrmion texture. Because the longitudinal ($b_1$) and transverse ($b_2$) magnetostrictive terms couple nonlinearly with the complex spatial gradients of the skyrmion magnetization, an exact analytical integration is intractable. We therefore evaluate this integral numerically in our micromagnetic simulations. The explicit tensor expansion of the force density and the reduction to a two-dimensional thin-film geometry are detailed in Appendix~\ref{sec:thiele}.

While the macroscopic integration accurately captures the skyrmion trajectory, it obscures the microscopic origin of the polarity-selective trapping. To reveal this, we decompose the magnetization into a static skyrmion profile $\mathbf{m}_0(\mathbf{r})$ and a dynamic linear response $\delta\mathbf{m}(\mathbf{r}, t)$ induced by the acoustic wave. A rigorous second-order perturbative expansion of the time-averaged force $\langle \mathbf{F}_{\text{mec}} \rangle_T$ (detailed in Appendix~\ref{sec:radiation}) strictly separates the interaction into two distinct components: $\langle \mathbf{F}_{\text{mec}} \rangle_T = \mathbf{F}_{\text{grad}} + \mathbf{F}_{\text{rad}}$. Here, the conservative gradient force $\mathbf{F}_{\text{grad}}$ originates from the reactive susceptibility $\boldsymbol{\chi}'$, whereas the non-conservative radiation force $\mathbf{F}_{\text{rad}}$ arises from the dissipative susceptibility $\boldsymbol{\chi}''$ via the time-averaged cross-correlation $\langle\delta\mathbf{m}\cdot\nabla\mathbf{H}_{\text{mec}}\rangle_T$.

In the present off-resonant driving regime, the gradient force is negligibly small ($\mathbf{F}_{\text{grad}} \approx 0$). The radiation force $\mathbf{F}_{\text{rad}}$, which intrinsically relies on the acoustic phonon spin texture, is directly linked to the dissipation landscape of Eq.~(\ref{eq:rayleigh}):
\begin{equation}
    \mathbf{F}_{\text{rad}} \;\propto\; -\,\nabla_{\!\mathbf{R}}\langle\dot{U}\rangle_T \;\propto\; -\,Q\;\nabla_{\!\mathbf{R}}\,S_{\text{p},z}(\mathbf{R}).
    \label{eq:force-dissipation}
\end{equation}
The first proportionality holds in the small-deformation regime ($\mathcal{M}\approx\mathcal{M}_0$); the second follows because $\langle\dot{U}\rangle_T$ inherits its spatial parity from $S_{\text{p},z}$. Equation~(\ref{eq:force-dissipation}) predicts that $Q=-1$ skyrmions are driven toward the global maximum of $\langle\dot{U}\rangle_T$ [Fig.~\ref{fig_skyrmion:fig2}(c)], while reversing $Q$ inverts the trapping direction. The gradient force $\mathbf{F}_{\text{grad}}$, being independent of $Q$, merely shifts the equilibrium uniformly without affecting the phonon-spin-mediated selective trapping.

The skyrmion mass $\mathcal{M}$, extracted by fitting micromagnetic simulation trajectories to Eq.~(\ref{Thiele Equation}), exhibits a strong position dependence as shown in Fig.~\ref{fig_skyrmion:fig2}(a). The mass reaches a local minimum near the attractive line, indicating that the skyrmion experiences minimal deformation and remains in a maximally relaxed state at this equilibrium position. In contrast, the mass increases significantly near the repulsive line and spin extrema where deformation is most pronounced.

The positions of the attractive and repulsive lines are intrinsically wavelength-dependent, as demonstrated in Fig.~\ref{fig_skyrmion:fig2}(d). In the point-particle limit $R_{\text{sk}}\ll\delta$, the trapping position coincides with the phonon spin maximum at $y=\delta/2$. For finite $R_{\text{sk}}/\delta$, the skyrmion samples the phonon spin landscape over its internal extent, introducing magnetization-dependent corrections that shift the equilibrium. In the long-wavelength limit ($\lambda \gg R_{\text{sk}}$), the net force vanishes as the field becomes uniform across the skyrmion. On the other hand, in the short-wavelength limit ($\lambda \ll R_{\text{sk}}$), the transverse displacement is quenched. The optimal trapping window thus lies at $\lambda \sim R_{\text{sk}}$.

\begin{figure*}
  \includegraphics[width=0.9\textwidth]{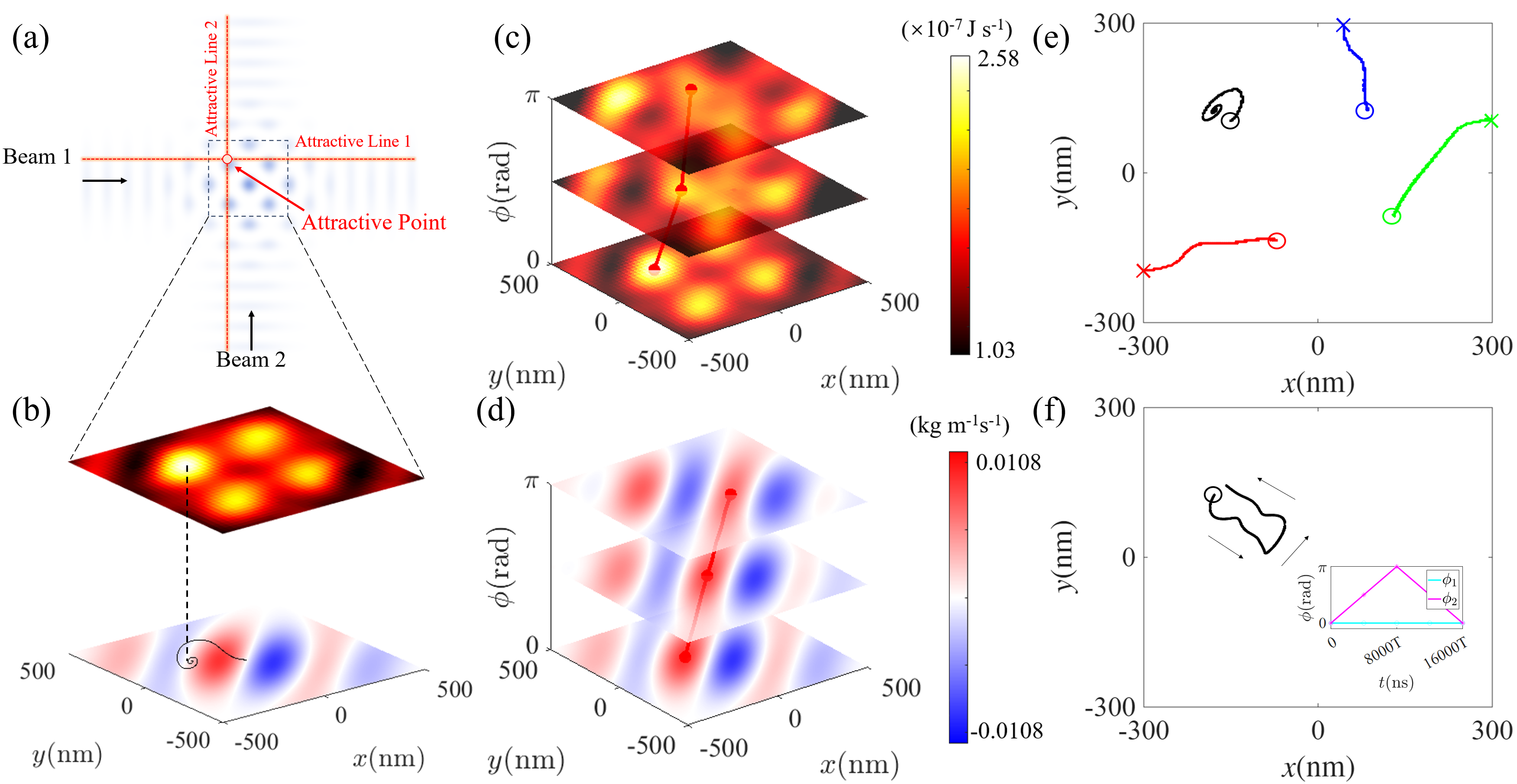}
  \centering
  \caption{\label{fig_skyrmion:fig3} Two-dimensional trapping and routing via orthogonal acoustic tweezers. The acoustic period is $T = \lambda/c_{\text{l}} \approx 83.2\,\text{ps}$. (a) Schematic of the tweezer architecture employing two orthogonally propagating Gaussian beams ($\lambda = 600\,\text{nm}$, $\delta = 400\,\text{nm}$) along $+\hat{\mathbf{x}}$ and $+\hat{\mathbf{y}}$ directions. Each beam possesses its own attractive line (red); the two attractive lines intersect to form an attractive point (red dot) that serves as the stable trapping site. (b) Zoomed view of the beam overlap region showing the Rayleigh dissipation rate (upper panel) and phonon spin distribution (lower panel). A skyrmion initialized at the origin migrates to and stabilizes at the attractive point over 7600 acoustic periods. (c) Spatiotemporal evolution of dissipation rate maxima (red dots) under quasi-static phase modulation of one beam from $\phi=0$ to $\phi=\pi$. (d) Corresponding evolution of the phonon spin spatial distribution, with the attractive point tracking the spin maximum. (e) Single-skyrmion selectivity in a multi-skyrmion ensemble simulated over 8000 acoustic periods: only the target skyrmion nearest the attractive point is captured while neighboring skyrmions are expelled. (f) Closed-loop precise routing of a trapped skyrmion along a programmable trajectory achieved by cyclic phase modulation ($0 \to \pi \to 0$) of one beam while maintaining constant phase for the orthogonal beam, with the total evolution spanning 16000 acoustic periods.}
\end{figure*}

\textit{Quasi-static trapping and routing via acoustic tweezers}--- A single Gaussian longitudinal acoustic beam creates an attractive line and a repulsive line parallel to its propagation axis, constraining skyrmion displacement only in the transverse ($y$) direction while leaving the longitudinal ($x$) position unconstrained. To achieve full two-dimensional trapping, we superimpose two orthogonally propagating Gaussian beams, a configuration realizable via orthogonal thin-film piezoelectric transducers \cite{Fu2021Engineering}. As schematically illustrated in Fig.~\ref{fig_skyrmion:fig3}(a), one beam propagates along the $+\hat{\mathbf{x}}$ direction while the other propagates along $+\hat{\mathbf{y}}$, causing their respective attractive lines to intersect at a well-defined attractive point. This intersection represents a stable equilibrium where skyrmions can be deterministically captured and held.

As shown in Fig.~\ref{fig_skyrmion:fig3}(b), a skyrmion initialized at the origin migrates quasi-statically to the attractive point, which coincides with the joint maximum of the Rayleigh dissipation rate and the phonon spin magnitude. The attractive point is reconfigurable: quasi-static phase modulation of one beam from $\phi=0$ to $\pi$ translates the intersection of the two attractive lines continuously through space [Figs.~\ref{fig_skyrmion:fig3}(c,d)]. In a multi-skyrmion ensemble [Fig.~\ref{fig_skyrmion:fig3}(e)], only the skyrmion nearest the attractive point is captured while all others are expelled by the surrounding repulsive force landscape. Closed-loop routing is demonstrated in Fig.~\ref{fig_skyrmion:fig3}(f) by cycling the phase $0\to\pi\to 0$, with the skyrmion tracking the moving attractive point with sub-nanometer precision along arbitrary programmable paths.

\textit{Discussion}--- The proposed acoustic tweezers offer distinct advantages over existing paradigms. Unlike localized field gradients or nanoscale current injectors that address individual skyrmions at the cost of complex nanofabrication and limited reconfigurability, our approach achieves single-particle selectivity intrinsically through the topological spin-texture of a global acoustic field, bypassing the need for physical nano-confinement. (i) Unlike optical tweezers that require high numerical aperture focusing incompatible with planar geometries, our planar acoustic approach operates entirely within the film plane via thin-film piezoelectric transducers, enabling straightforward on-chip integration. (ii) The mechanism circumvents the Joule heating and electromigration inherent to spin-polarized current driving, ensuring non-destructive manipulation that preserves the topological integrity of the skyrmion. (iii) The trapping relies on phonon-spin-mediated selective interaction rather than direct momentum transfer, enabling polarity-selective control unattainable by conventional global fields. (iv) The principal limitation concerns the requirement $\lambda \sim \delta$ for significant phonon spin generation, restricting the manipulation range to the focal region; however, this confinement complements macroscopic surface acoustic wave transport, suggesting hierarchical protocols combining global ensemble translation with deterministic single-skyrmion trapping. (v) The quasi-static, one-way coupling approximation is justified because the skyrmion drift velocity ($\sim\text{m/s}$) is many orders of magnitude below the sound velocity ($\sim\text{km/s}$), and the driving frequency remains off-resonant with respect to the skyrmion eigenmodes. The adopted frequency $f = c_{\text{l}}/\lambda \approx 12.0\,\text{GHz}$ falls within the operational range of state-of-the-art film bulk acoustic resonators. Magnetic dipole-dipole interactions are negligible for inter-skyrmion separations exceeding $3R_{\text{sk}} \approx 80\,\text{nm}$. Furthermore, the parity-locked interaction mechanism established here shares a profound mathematical isomorphism with the transverse spin of evanescent optical waves, suggesting a universal paradigm for the topological sorting and manipulation of diverse chiral quasiparticles, such as magnetic vortices and chiral domain walls, via classical wave fields.

\textit{Conclusion}--- We have established a paradigm for magnetic skyrmion manipulation that bridges the conceptual gap between global field driving and individual particle control. By revealing that spatially confined Gaussian longitudinal acoustic beams carry nontrivial phonon spin angular momentum with domains of opposite polarity, we demonstrated how magnetoelastic coupling creates an effective magnetic field whose parity structure is locked to the acoustic spin texture. This parity correspondence enables polarity-selective radiation forces that attract skyrmions toward phonon spin maxima while repelling them from minima, creating attractive and repulsive lines in the acoustic field. The strategic intersection of two orthogonally propagating beams transforms these lines into a reconfigurable attractive point that captures individual skyrmions with sub-nanometer precision. This global-field-local-interaction scheme provides a viable, non-destructive route for deterministic single-skyrmion control within dense topological ensembles through phonon-spin-mediated selective trapping, establishing a solid foundation for programmable spintronic architectures.

\textit{Acknowledgments}--- This work was supported by National Key Research Program of China (Grant No. 2022YFA1403300), the Innovation Program for Quantum Science and Technology (Grant No. 2024ZD0300103), and National Natural Science Foundation of China (Grant No. 12574112 and 12204107).

\bibliographystyle{apsrev4-2}
\bibliography{paper_skyrmion}

\clearpage
\appendix
\onecolumngrid

\section{DERIVATION OF THE PHONON-SPIN-INDUCED MAGNETOELASTIC CHIRALITY}
\label{sec:chirality}

\subsection{Kinematic Origin of the Transverse Displacement}
In an isotropic elastic medium, a pure longitudinal wave must satisfy the irrotational condition $\nabla \times \mathbf{u} = 0$. In the 2D $x$-$y$ plane, this mandates:
\begin{equation}
    \frac{\partial u_y}{\partial x} - \frac{\partial u_x}{\partial y} = 0 \implies \frac{\partial u_y}{\partial x} = \frac{\partial u_x}{\partial y}.
\end{equation}
Assume the longitudinal displacement is a plane wave with a transverse Gaussian envelope:
\begin{equation}
    u_x(x,y,t) = u_0 e^{-y^2/\delta^2} \cos(kx - \omega t) \equiv A(y) \cos\Phi,
\end{equation}
where $\Phi = kx - \omega t$ and $A(y) = u_0 e^{-y^2/\delta^2}$. Taking the partial derivative with respect to $y$:
\begin{equation}
    \frac{\partial u_x}{\partial y} = -u_0 \frac{2y}{\delta^2} e^{-y^2/\delta^2} \cos(kx - \omega t).
\end{equation}
Integrating $\partial u_y / \partial x = \partial u_x / \partial y$ with respect to $x$ yields the exact transverse displacement:
\begin{equation}
    u_y(x,y,t) = \int \left( -u_0 \frac{2y}{\delta^2} e^{-y^2/\delta^2} \cos(kx - \omega t) \right) dx = -u_0 \frac{2y}{k\delta^2} e^{-y^2/\delta^2} \sin(kx - \omega t) \equiv -B(y) \sin\Phi.
\end{equation}
The integration over $x$ inherently shifts the phase by $\pi/2$ ($\cos \to \sin$), while the derivative over $y$ extracts the odd spatial parity factor $y$. This mathematically guarantees that the local lattice vibration is elliptically polarized, and the handedness reverses across the beam axis ($y=0$).

\subsection{Phonon Spin Density and Spatial Parity}
The phonon spin angular momentum density is defined as: $\mathbf{S}_{\text{p}} = \rho \langle \mathbf{u} \times \dot{\mathbf{u}} \rangle$.
Calculating the time derivatives ($\dot{\Phi} = -\omega$):
\begin{equation}
    \dot{u}_x = A(y)\, \omega \sin\Phi, \quad \dot{u}_y = B(y)\, \omega \cos\Phi.
\end{equation}
The $z$-component of the phonon spin is evaluated via the cross product:
\begin{align}
    S_{\text{p},z} &= \rho \langle u_x \dot{u}_y - u_y \dot{u}_x \rangle \notag \\
    &= \rho \langle (A \cos\Phi)(B \omega \cos\Phi) - (-B \sin\Phi)(A \omega \sin\Phi) \rangle \notag \\
    &= \rho \omega A(y) B(y) \langle \cos^2\Phi + \sin^2\Phi \rangle.
\end{align}
Since $\langle \cos^2\Phi + \sin^2\Phi \rangle = 1$, the time-averaging yields:
\begin{equation}
    S_{\text{p},z} = \rho \omega A(y) B(y) = \frac{2\rho u_0^2 c_{\text{l}} y}{\delta^2} \exp\left(-\frac{2y^2}{\delta^2}\right),
\end{equation}
where $c_{\text{l}} = \omega/k$. The phonon spin density $S_{\text{p},z}$ is an odd function of $y$ ($S_{\text{p},z} \propto y$), demonstrating that the phonon spin domains are antisymmetric across the beam axis.

\subsection{Strain Tensor and Magnetoelastic Effective Field}
The coupling to the magnetic system is mediated by the strain tensor $\varepsilon_{ij} = \frac{1}{2}(\partial_i u_j + \partial_j u_i)$. The dominant in-plane components are:
\begin{align}
    \varepsilon_{xx} &= \partial_x u_x = -k A(y) \sin\Phi, \\
    \varepsilon_{yy} &= \partial_y u_y = -B'(y) \sin\Phi, \\
    \varepsilon_{xy} &= \frac{1}{2}(\partial_y u_x + \partial_x u_y) = \partial_y u_x = A'(y) \cos\Phi \equiv -C(y) \cos\Phi,
\end{align}
where $B'(y) = \frac{2u_0}{k\delta^2} \left( 1 - \frac{2y^2}{\delta^2} \right) e^{-y^2/\delta^2}$ and $C(y) = -A'(y) = k B(y) = u_0 \frac{2y}{\delta^2} e^{-y^2/\delta^2}$.

Based on the magnetoelastic energy functional for a cubic lattice, the in-plane effective magnetic field $\mathbf{H}_{\text{mec}}$ is derived via $\mathbf{H}_{\text{mec}} = -\frac{1}{\mu_0 M_s} \frac{\partial \mathcal{E}_{\text{me}}}{\partial \mathbf{m}}$:
\begin{align}
    H_{x}^{\text{mec}} &= -\frac{2}{\mu_0 M_s} (b_1 \varepsilon_{xx} m_x + b_2 \varepsilon_{xy} m_y) = X \sin\Phi + Y \cos\Phi, \\
    H_{y}^{\text{mec}} &= -\frac{2}{\mu_0 M_s} (b_1 \varepsilon_{yy} m_y + b_2 \varepsilon_{xy} m_x) = Z \sin\Phi + W \cos\Phi,
\end{align}
where the spatially dependent coefficients are explicitly defined as:
\begin{equation}
    X = \frac{2b_1 k A}{\mu_0 M_s} m_x, \quad Y = \frac{2b_2 C}{\mu_0 M_s} m_y, \quad Z = \frac{2b_1 B'}{\mu_0 M_s} m_y, \quad W = \frac{2b_2 C}{\mu_0 M_s} m_x.
\end{equation}

\subsection{Field Chirality Density and Parity Correspondence}
We define the field chirality density analogously to angular momentum:
\begin{equation}
    \mathcal{C}_H = \langle H_{x}^{\text{mec}} \dot{H}_{y}^{\text{mec}} - H_{y}^{\text{mec}} \dot{H}_{x}^{\text{mec}} \rangle.
\end{equation}
Substituting the time derivatives $\dot{H}_{x}^{\text{mec}} = \omega(-X \cos\Phi + Y \sin\Phi)$ and $\dot{H}_{y}^{\text{mec}} = \omega(-Z \cos\Phi + W \sin\Phi)$ into the cross product:
\begin{align}
    H_{x}^{\text{mec}} \dot{H}_{y}^{\text{mec}} - H_{y}^{\text{mec}} \dot{H}_{x}^{\text{mec}} &= \omega \big[ (XW - YZ)\sin^2\Phi + (XW - YZ)\cos^2\Phi \big] = \omega (XW - YZ).
\end{align}
Since the result is strictly independent of time, the time-averaging yields the exact analytical expression:
\begin{equation}
    \mathcal{C}_H = \omega (XW - YZ) = \frac{4\omega b_1 b_2 C(y)}{(\mu_0 M_s)^2} \Big[ k A(y) m_x^2 - B'(y) m_y^2 \Big].
\end{equation}
Comparing $S_{\text{p},z} \propto A(y)B(y) \propto y$ with $\mathcal{C}_H \propto C(y)[\dots] \propto y$, we observe that both the macroscopic phonon spin and the local effective field chirality share the exact same spatial odd-parity sign factor, $\text{Sgn}(y)$. This mathematically proves the parity correspondence between the phonon spin and the magnetoelastic field chirality.

\section{GENERALIZED THIELE EQUATION AND INERTIAL DYNAMICS WITH MAGNETOELASTIC COUPLING}
\label{sec:thiele}

To describe the dynamics of a skyrmion, the translational motion of the magnetization texture must be investigated. The dynamics are dominated by the LLG equation, which can be extended to Thiele's model \cite{PhysRevLett.30.230,Mochizuki2014}.
\begin{align}
  \frac{\partial \mathbf{m}}{\partial t}=-\gamma\mathbf{m}\times\mathbf{H}_{\text{eff}}+\alpha\mathbf{m}\times\frac{\partial\mathbf{m}}{\partial t}.
\end{align}
Using the vector identity $\mathbf{m}\times(\mathbf{m}\times\partial_t\mathbf{m})=(\mathbf{m}\cdot\partial_t\mathbf{m})\mathbf{m}-(\mathbf{m}\cdot\mathbf{m})\partial_t\mathbf{m}=-\partial_t\mathbf{m}$, the LLG equation can be rewritten in a compact form:
\begin{align}
  \mathbf{m}\times(\mathbf{H}^{\text{g}}+\mathbf{H}^{\text{e}}+\mathbf{H}^{\text{d}})=0,
\end{align}
where $\mathbf{H}^{\text{g}}=-\frac{1}{\gamma}\mathbf{m}\times\partial_t\mathbf{m}$ is the gyroscopic term, $\mathbf{H}^{\text{e}}=\mathbf{H}_{\text{eff}}$ is the external term, and $\mathbf{H}^{\text{d}}=-\frac{\alpha}{\gamma}\partial_t\mathbf{m}$ is the damping term. Note that all terms here have the dimension of effective magnetic field (A/m).

The force density is defined by the energy density of the magnetization:
\begin{align}
  f_i = -\gamma\sum_j \frac{\partial m_j}{\partial x_i} H_j,
\end{align}
where $\mathbf{H}=\mathbf{H}^{\text{g}}+\mathbf{H}^{\text{e}}+\mathbf{H}^{\text{d}}$. In a steady translational state, the total force density must be in balance:
\begin{align}
  \mathbf{f}^{\text{g}}+\mathbf{f}^{\text{d}}+\mathbf{f}^{\text{e}}=0.
\end{align}

The force density must be expanded to reveal its physical meaning. For a moving magnetic texture $\mathbf{m}(\mathbf{r},t)$, the time derivative is transformed using the texture velocity $\mathbf{v}$:
\begin{align}
  \frac{\partial \mathbf{m}}{\partial t} &= \frac{\partial \mathbf{m}}{\partial x}\frac{\partial x}{\partial t}+\frac{\partial \mathbf{m}}{\partial y}\frac{\partial y}{\partial t}+\frac{\partial \mathbf{m}}{\partial z}\frac{\partial z}{\partial t} \notag \\
    &= -\left(
    \begin{array}{c}
      v_x\frac{\partial m_x}{\partial x}+v_y\frac{\partial m_x}{\partial y}+v_z\frac{\partial m_x}{\partial z} \\
      v_x\frac{\partial m_y}{\partial x}+v_y\frac{\partial m_y}{\partial y}+v_z\frac{\partial m_y}{\partial z} \\
      v_x\frac{\partial m_z}{\partial x}+v_y\frac{\partial m_z}{\partial y}+v_z\frac{\partial m_z}{\partial z} \\
    \end{array}
    \right).
\end{align}
The negative sign originates from the convective derivative of the moving texture $\mathbf{m}(\mathbf{r}-\mathbf{v}t,t)$. Expanding the gyroscopic term yields:
\begin{align}
  f^{\text{g}}_x &= -\gamma\left(\frac{\partial m_x}{\partial x}H^{\text{g}}_x+\frac{\partial m_y}{\partial x}H^{\text{g}}_y+\frac{\partial m_z}{\partial x}H^{\text{g}}_z\right) \notag \\
    &= \mathbf{m}\cdot\left(\frac{\partial \mathbf{m}}{\partial x}\times\frac{\partial \mathbf{m}}{\partial y}\right)v_y+\mathbf{m}\cdot\left(\frac{\partial \mathbf{m}}{\partial x}\times\frac{\partial \mathbf{m}}{\partial z}\right)v_z.
\end{align}
Following the same derivation, the other components of the force density are:
\begin{align}
    f_y^{\text{g}} &= \mathbf{m}\cdot\left(\frac{\partial \mathbf{m}}{\partial y}\times\frac{\partial \mathbf{m}}{\partial x}\right)v_x+\mathbf{m}\cdot\left(\frac{\partial \mathbf{m}}{\partial y}\times\frac{\partial \mathbf{m}}{\partial z}\right)v_z, \notag \\
    f_z^{\text{g}} &= \mathbf{m}\cdot\left(\frac{\partial \mathbf{m}}{\partial z}\times\frac{\partial \mathbf{m}}{\partial x}\right)v_x+\mathbf{m}\cdot\left(\frac{\partial \mathbf{m}}{\partial z}\times\frac{\partial \mathbf{m}}{\partial y}\right)v_y.
\end{align}
The gyroscopic force can be expressed as $\mathbf{f}^{\text{g}}=\mathbf{g}\times\mathbf{v}$. The gyrocoupling vector is:
\begin{align}
  \mathbf{g}=-
  \left(
  \begin{array}{c}
    \mathbf{m}\cdot\left(\frac{\partial\mathbf{m}}{\partial y}\times\frac{\partial\mathbf{m}}{\partial z}\right) \\
    \mathbf{m}\cdot\left(\frac{\partial\mathbf{m}}{\partial z}\times\frac{\partial\mathbf{m}}{\partial x}\right) \\
    \mathbf{m}\cdot\left(\frac{\partial\mathbf{m}}{\partial x}\times\frac{\partial\mathbf{m}}{\partial y}\right) \\
  \end{array}
  \right).
\end{align}
For a skyrmion in the $xy$ plane, the total gyroscopic force is the integral of the force density over the whole space:
\begin{align}
    \mathbf{G} &= \int_{\text{UC}}\mathrm{d}^2r(-\mathbf{g}) =
    \left(
    \begin{array}{c}
        0 \\ 0 \\ 4\pi Q
     \end{array}
    \right).
\end{align}
The total force is $\mathbf{F}^{\text{g}}=\mathbf{G}\times(-\mathbf{v})$, where $Q=\frac{1}{4\pi}\int_{\text{UC}}\mathrm{d}^2r\mathbf{m}\cdot\left(\frac{\partial\mathbf{m}}{\partial x}\times\frac{\partial\mathbf{m}}{\partial y}\right)=-1$ is the topological winding number. $\mathbf{F}^{\text{g}}$ is commonly referred to as the Magnus force.

The damping term is expanded as:
\begin{align}
  f^{\text{d}}_x &= -\gamma\left(H_x^{\text{d}}\frac{\partial m_x}{\partial x}+H_y^{\text{d}}\frac{\partial m_y}{\partial x}+H_z^{\text{d}}\frac{\partial m_z}{\partial x}\right) \notag \\
    &= -\alpha\left(\frac{\partial m_x}{\partial x}\frac{\partial m_x}{\partial x}+\frac{\partial m_y}{\partial x}\frac{\partial m_y}{\partial x}+\frac{\partial m_z}{\partial x}\frac{\partial m_z}{\partial x}\right)v_x \notag \\
    &\quad -\alpha\left(\frac{\partial m_x}{\partial x}\frac{\partial m_x}{\partial y}+\frac{\partial m_y}{\partial x}\frac{\partial m_y}{\partial y}+\frac{\partial m_z}{\partial x}\frac{\partial m_z}{\partial y}\right)v_y \notag \\
    &\quad -\alpha\left(\frac{\partial m_x}{\partial x}\frac{\partial m_x}{\partial z}+\frac{\partial m_y}{\partial x}\frac{\partial m_y}{\partial z}+\frac{\partial m_z}{\partial x}\frac{\partial m_z}{\partial z}\right)v_z.
\end{align}
The damping force can be expressed as $\mathbf{F}^{\text{d}}=-\alpha\overset{\leftrightarrow}{\mathbf{D}}\cdot\mathbf{v}$, where the damping tensor is:
\begin{align}
  \overset{\leftrightarrow}{\mathbf{D}}_{ij}=\int_{\text{UC}}\mathrm{d}^2r\frac{\partial\mathbf{m}}{\partial x_i}\frac{\partial\mathbf{m}}{\partial x_j}.
\end{align}

Considering a cubic crystal with magnetoelastic coupling, the magnetoelastic energy density is given by $\mathcal{E}_{\text{me}} = b_1 \sum_i \varepsilon_{ii} m_i^2 + b_2 \sum_{i \neq j} \varepsilon_{ij} m_i m_j$, where $\varepsilon_{ij} = \frac{1}{2}(\partial_i u_j + \partial_j u_i)$ is the strain tensor. The effective magnetic field arising from magnetoelastic coupling, $\mathbf{H}_{\text{mec}} = -\frac{1}{\mu_0 M_s}\frac{\partial \mathcal{E}_{\text{me}}}{\partial \mathbf{m}}$, can be decomposed into longitudinal ($b_1$) and transverse ($b_2$) components:
\begin{align}
    \mathbf{H}^{b_1} &= -\frac{2b_1}{\mu_0 M_s}
    \begin{pmatrix}
        m_x \partial_x u_x \\
        m_y \partial_y u_y \\
        m_z \partial_z u_z
    \end{pmatrix}, \label{eq:Hb1} \\
    \mathbf{H}^{b_2} &= -\frac{b_2}{\mu_0 M_s}
    \begin{pmatrix}
        m_y (\partial_x u_y + \partial_y u_x) + m_z (\partial_x u_z + \partial_z u_x) \\
        m_x (\partial_x u_y + \partial_y u_x) + m_z (\partial_y u_z + \partial_z u_y) \\
        m_x (\partial_x u_z + \partial_z u_x) + m_y (\partial_y u_z + \partial_z u_y)
    \end{pmatrix}. \label{eq:Hb2}
\end{align}
Substituting these effective fields into the force density definition $f_i = -\gamma \sum_j \frac{\partial m_j}{\partial x_i} H_{\text{mec}, j}$, we obtain the longitudinal displacement field force density term $\mathbf{f}^{b_1}$:
\begin{equation}
    \mathbf{f}^{b_1} = \frac{2\gamma b_1}{\mu_0 M_s}
    \begin{pmatrix}
        m_x \partial_x u_x \partial_x m_x + m_y \partial_y u_y \partial_x m_y + m_z \partial_z u_z \partial_x m_z \\
        m_x \partial_x u_x \partial_y m_x + m_y \partial_y u_y \partial_y m_y + m_z \partial_z u_z \partial_y m_z \\
        m_x \partial_x u_x \partial_z m_x + m_y \partial_y u_y \partial_z m_y + m_z \partial_z u_z \partial_z m_z
    \end{pmatrix}.
    \label{eq:fb1}
\end{equation}
Similarly, the transverse displacement field force density term $\mathbf{f}^{b_2}$ is derived as:
\begin{align}
    f^{b_2}_x &= \frac{\gamma b_2}{\mu_0 M_s} \Big[ (m_y \partial_x m_x + m_x \partial_x m_y)(\partial_x u_y + \partial_y u_x) \notag \\
    &\quad + (m_z \partial_x m_x + m_x \partial_x m_z)(\partial_x u_z + \partial_z u_x) \notag \\
    &\quad + (m_z \partial_x m_y + m_y \partial_x m_z)(\partial_y u_z + \partial_z u_y) \Big], \notag \\
    f^{b_2}_y &= \frac{\gamma b_2}{\mu_0 M_s} \Big[ (m_y \partial_y m_x + m_x \partial_y m_y)(\partial_x u_y + \partial_y u_x) \notag \\
    &\quad + (m_z \partial_y m_x + m_x \partial_y m_z)(\partial_x u_z + \partial_z u_x) \notag \\
    &\quad + (m_z \partial_y m_y + m_y \partial_y m_z)(\partial_y u_z + \partial_z u_y) \Big], \notag \\
    f^{b_2}_z &= \frac{\gamma b_2}{\mu_0 M_s} \Big[ (m_y \partial_z m_x + m_x \partial_z m_y)(\partial_x u_y + \partial_y u_x) \notag \\
    &\quad + (m_z \partial_z m_x + m_x \partial_z m_z)(\partial_x u_z + \partial_z u_x) \notag \\
    &\quad + (m_z \partial_z m_y + m_y \partial_z m_z)(\partial_y u_z + \partial_z u_y) \Big].
    \label{eq:fb2}
\end{align}
For a 2D magnetic thin film system in the $x$-$y$ plane, terms containing spatial derivatives with respect to $z$ (e.g., $\partial_z m_i$ and $\partial_z u_i$) vanish or are negligible. The total magnetoelastic coupling force acting on the skyrmion is obtained by integrating the force density over the simulation volume. Since the analytical integration of the highly nonlinear cross-terms between the skyrmion magnetization texture and the spatially varying acoustic strain field is intractable, the total force is evaluated numerically:
\begin{align}
  \mathbf{F}^{\text{mec}}=\int_{\text{UC}}\mathrm{d}^2r(\mathbf{f}^{b_1}+\mathbf{f}^{b_2})=\mathbf{G}\times\mathbf{v}+\alpha\overset{\leftrightarrow}{\mathbf{D}}\cdot\mathbf{v}.
  \label{eq:Fmec_integral}
\end{align}

To accurately model the observed trajectory, it is imperative to employ a generalized Thiele equation that accounts for substantial non-equilibrium skyrmion deformations under large-amplitude acoustic excitation. The breakdown of the standard Thiele formalism originates from field-induced deviations from rigid-body skyrmion topology, which introduce an inertial mass term $\mathcal{M}$:
\begin{align}
  \mathbf{F}_{\text{mec}}=\mathbf{G}\times\mathbf{v}+\alpha\overset{\leftrightarrow}{\mathbf{D}}\cdot\mathbf{v}+\mathcal{M}\dot{\mathbf{v}}.
\end{align}
Expressing the generalized Thiele equation in Cartesian components yields the coupled dynamical system:
\begin{align}
    4\pi v_y+\alpha D_{11}v_x+\mathcal{M}\frac{\mathrm{d}v_x}{\mathrm{d}t} &= F_x, \notag \\
    -4\pi v_x+\alpha D_{22}v_y+\mathcal{M}\frac{\mathrm{d}v_y}{\mathrm{d}t} &= F_y.
\end{align}
The effective mass $\mathcal{M}$ exhibits a linear proportionality to the deformation magnitude $\Delta\mathbf{m}=\mathbf{m}-\mathbf{m}_0$. Discretizing the time derivative to solve for the velocity requires the initial velocity $\mathbf{v}_0$:
\begin{align}
    4\pi v_y\Delta t+\alpha D_{11}v_x\Delta t+\mathcal{M}(v_x-v_{x0}) &= F_x\Delta t, \notag \\
    -4\pi v_x\Delta t+\alpha D_{22}v_y\Delta t+\mathcal{M}(v_y-v_{y0}) &= F_y\Delta t.
\end{align}
The matrix form of these linear equations is:
\begin{align}
      &\left(
      \begin{array}{cc}
        \alpha D_{11}\Delta t+\mathcal{M} & 4\pi \Delta t \\
        -4\pi \Delta t & \alpha D_{22}\Delta t+\mathcal{M}
      \end{array}
      \right)
      \left(
      \begin{array}{c}
        v_x \\
        v_y
      \end{array}
      \right) =
      \left(
      \begin{array}{c}
        F_x\Delta t+\mathcal{M}v_{x0} \\
        F_y\Delta t+\mathcal{M}v_{y0}
      \end{array}
      \right).
\end{align}
Solving this system yields the explicit velocity components:
\begin{align}
    v_x &= \frac{(\alpha D_{22}\Delta t+\mathcal{M})(F_x\Delta t+\mathcal{M}v_{x0}) - 4\pi \Delta t(F_y\Delta t+\mathcal{M}v_{y0})}{(\alpha D_{11}\Delta t+\mathcal{M})(\alpha D_{22}\Delta t+\mathcal{M})+(4\pi \Delta t)^2}, \notag \\
    v_y &= \frac{(\alpha D_{11}\Delta t+\mathcal{M})(F_y\Delta t+\mathcal{M}v_{y0}) + 4\pi \Delta t(F_x\Delta t+\mathcal{M}v_{x0})}{(\alpha D_{11}\Delta t+\mathcal{M})(\alpha D_{22}\Delta t+\mathcal{M})+(4\pi \Delta t)^2}.
\end{align}

By setting $\mathcal{M} = 0$, the governing equations reduce to the steady-state motion for undeformed skyrmions:
\begin{align}
    v_x &= \frac{\alpha D_{22}F_x}{\alpha^2D_{11}D_{22}+(4\pi)^2}, \notag \\
    v_y &= \frac{\alpha D_{11}F_y}{\alpha^2D_{11}D_{22}+(4\pi)^2}.
\end{align}
The inclusion of inertial mass $\mathcal{M}$ establishes a Newtonian-like dependence on initial velocity $\mathbf{v}_0$, transforming the skyrmion dynamics from a purely gyroscopic system to quasi-Newtonian particle behavior.

For skyrmions subjected to static displacement fields, spatial field gradients induce effective driving forces. Because the displacement field is time-independent, the resulting dynamics are governed entirely by the conservative gradient force $\mathbf{F}_{\text{grad}}$; the radiation force $\mathbf{F}_{\text{rad}}$, which requires finite temporal oscillation to generate a non-zero second-order cross-correlation, is identically zero in this static limit. The magnetoelastic coupling generates force components that decompose into a topological Magnus contribution $\mathbf{v}_{\text{g}}$ and a damping drag component $\mathbf{v}_{\text{d}}$, with resultant trajectory $\mathbf{v}=\mathbf{v}_{\text{g}}+\mathbf{v}_{\text{d}}$. The velocity ratio $v_{\text{g}}/v_{\text{d}}$ decreases with increasing $\alpha$.

These findings demonstrate that a skyrmion with a larger effective mass tends to maintain its original state of motion more effectively. Deviations from the standard Thiele equation in certain regions of the Gaussian acoustic field suggest that local deformations of the skyrmion may induce variations in its effective mass. To quantitatively describe the altered trajectories of skyrmions in Gaussian acoustic fields, we propose fitting the trajectories by estimating the mass. Specifically, we constrain the skyrmion mass to ensure that, after 200 oscillation cycles, the terminal velocities $\mathbf{v}$ are equal to the velocities calculated by simulation. Under this condition, the effective mass $\mathcal{M}$ can be extracted by solving the following quadratic equations:
\begin{align}
\begin{split}
    (v_{x}-v_{x0})\mathcal{M}^2 &+ (\alpha D_{11}v_{x}+\alpha D_{22}v_{x}-\alpha D_{22}v_{x0}-F_x+4\pi v_{y0})\Delta t\,\mathcal{M} \\
    &+ (\alpha^2D_{11}D_{22}v_{x}+16\pi^2v_{x}-\alpha D_{22}F_x+4\pi F_y)(\Delta t)^2=0,
\end{split} \label{eq:mass_quadratic_x} \\
\begin{split}
    (v_{y}-v_{y0})\mathcal{M}^2 &+ (\alpha D_{11}v_{y}+\alpha D_{22}v_{y}-\alpha D_{11}v_{y0}-F_y-4\pi v_{x0})\Delta t\,\mathcal{M} \\
    &+ (\alpha^2D_{11}D_{22}v_{y}+16\pi^2v_{y}-\alpha D_{11}F_y-4\pi F_x)(\Delta t)^2=0.
\end{split} \label{eq:mass_quadratic_y}
\end{align}

\section{MICROSCOPIC DERIVATION OF THE MAGNETOELASTIC RADIATION FORCE}
\label{sec:radiation}

To expose the microscopic origin of the acoustic tweezing force and rigorously justify the non-zero time-averaged macroscopic force, we start from the instantaneous force density exerted by the magnetoelastic effective field on the magnetization texture. To maintain dimensional consistency with the generalized Thiele equation derived in Appendix~\ref{sec:thiele}, the force density is defined as:
\begin{equation}
    \mathbf{f}(\mathbf{r}, t) = -\gamma\sum_j (\nabla m_j) H_{\text{mec}, j}.
    \label{eq:f_density_C}
\end{equation}
We emphasize the explicit connection to Appendix~\ref{sec:thiele}: Eq.~(\ref{eq:f_density_C}) is obtained by restricting the general force-density definition $f_i=-\gamma\sum_j(\partial_i m_j)H_j$ to the magnetoelastic effective field $\mathbf{H}_{\text{mec}}=\mathbf{H}^{b_1}+\mathbf{H}^{b_2}$ [Eqs.~(\ref{eq:Hb1})--(\ref{eq:Hb2})]. It therefore coincides exactly with the magnetoelastic force density $\mathbf{f}^{\,\text{mec}}=\mathbf{f}^{\,b_1}+\mathbf{f}^{\,b_2}$ derived therein [Eqs.~(\ref{eq:fb1})--(\ref{eq:fb2})]. Consequently, the volume integral of $\mathbf{f}(\mathbf{r},t)$ is identical to the instantaneous magnetoelastic driving force $\mathbf{F}^{\text{mec}}(t)$ that enters the right-hand side of the generalized Thiele equation [Eq.~(\ref{eq:Fmec_integral})]. Utilizing integration by parts and assuming the magnetization perturbations vanish at the boundaries (justified by the localized nature of the skyrmion texture and the Gaussian decay of the acoustic field), the force can be recast into the form of a dipole-field gradient interaction:
\begin{equation}
    \mathbf{F}^{\text{mec}}(t) = \int \mathrm{d}^2 r \, \mathbf{f} = \gamma\int \mathrm{d}^2 r \sum_j m_j \nabla H_{\text{mec}, j}.
    \label{eq:Fmec_inst}
\end{equation}
The time average of Eq.~(\ref{eq:Fmec_inst}) over one acoustic period, $\langle\mathbf{F}^{\text{mec}}\rangle_T$, is precisely the $\mathbf{F}_{\text{mec}}$ governing the centre-of-mass dynamics in the generalized Thiele equation.

We decompose the magnetization into a static skyrmion profile $\mathbf{m}_0(\mathbf{r})$ and a dynamic linear response $\delta\mathbf{m}(\mathbf{r}, t)$ induced by the acoustic wave, with $|\delta\mathbf{m}| \ll 1$ and $\delta\mathbf{m} \perp \mathbf{m}_0$. The time-averaged force $\langle \mathbf{F}^{\text{mec}} \rangle_T$ from the static part $\mathbf{m}_0$ strictly vanishes because $\mathbf{H}_{\text{mec}}$ oscillates harmonically with zero mean. The non-zero DC force originates entirely from the dynamic cross-correlation:
\begin{equation}
    \langle \mathbf{F}^{\text{mec}} \rangle_T = \gamma\int \mathrm{d}^2 r \sum_j \langle \delta m_j \nabla H_{\text{mec}, j} \rangle_T.
\end{equation}

Under a harmonic acoustic drive, the magnetoelastic field takes the generic form:
\begin{equation}
    \mathbf{H}_{\text{mec}}(\mathbf{r}, t) = \mathbf{h}_c(\mathbf{r}) \cos(\omega t) + \mathbf{h}_s(\mathbf{r}) \sin(\omega t),
\end{equation}
where $\mathbf{h}_c$ and $\mathbf{h}_s$ are the spatially-dependent amplitude vectors given explicitly in Appendix~\ref{sec:chirality}. The local dynamic magnetization is governed by the complex susceptibility tensor $\boldsymbol{\chi}(\omega) = \boldsymbol{\chi}'(\omega) + i\boldsymbol{\chi}''(\omega)$, defined through $\delta m_j = \sum_k \chi_{jk} H_{\text{mec},k}$ so that $[\chi_{jk}] = \text{m/A}$, consistent with the dimensionless convention for $\mathbf{m}$. Adopting the $e^{-i\omega t}$ Fourier convention, the time-domain linear response reads:
\begin{equation}
    \delta m_j(\mathbf{r}, t) = \sum_k \left[ \chi'_{jk} H_{\text{mec}, k}(t) - \chi''_{jk} \frac{\dot{H}_{\text{mec}, k}(t)}{\omega} \right],
\end{equation}
where $\boldsymbol{\chi}'$ (reactive part) describes the in-phase, energy-storing response, and $\boldsymbol{\chi}''$ (dissipative part) describes the quadrature, energy-absorbing response. The minus sign before $\chi''$ follows from the $e^{-i\omega t}$ convention: for $H(t) = H_0\cos\omega t$, the dissipative response is $+\chi'' H_0 \sin\omega t = -\chi'' \dot{H}/\omega$.

We now compute the time-averaged force explicitly. The field gradient is:
\begin{equation}
    \nabla H_{\text{mec}, j}(t) = (\nabla h_{c,j}) \cos\omega t + (\nabla h_{s,j}) \sin\omega t,
\end{equation}
and the time derivative divided by $\omega$ is:
\begin{equation}
    \frac{\dot{H}_{\text{mec}, k}(t)}{\omega} = -h_{c,k} \sin\omega t + h_{s,k} \cos\omega t.
\end{equation}

Substituting into $\langle \delta m_j \nabla H_{\text{mec}, j} \rangle_T$ and using the orthogonality relations $\langle \cos^2\omega t \rangle = \langle \sin^2\omega t \rangle = 1/2$ and $\langle \cos\omega t \sin\omega t \rangle = 0$, we obtain two distinct force contributions:
\begin{equation}
    \langle \mathbf{F}^{\text{mec}} \rangle_T = \mathbf{F}_{\text{grad}} + \mathbf{F}_{\text{rad}}.
\end{equation}

\textbf{Gradient force.} The contribution from $\boldsymbol{\chi}'$ is:
\begin{equation}
    \mathbf{F}_{\text{grad}} = -\frac{\gamma}{4} \int \mathrm{d}^2 r \sum_{j,k} (\nabla \chi'_{jk}) \left( h_{c,k} h_{c,j} + h_{s,k} h_{s,j} \right).
    \label{eq:F_grad}
\end{equation}
This force arises from the spatial gradient of the local susceptibility coupled to the acoustic field intensity. It is non-chiral in nature and analogous to the conservative dipole force in optical tweezers.

\textbf{Radiation force.} The contribution from $\boldsymbol{\chi}''$ is:
\begin{align}
    \mathbf{F}_{\text{rad}} &= -\frac{\gamma}{2} \int \mathrm{d}^2 r \sum_{j,k} \chi''_{jk} \left( h_{s,k} \nabla h_{c,j} - h_{c,k} \nabla h_{s,j} \right).
    \label{eq:F_rad}
\end{align}
This is the radiation force (analogous to the scattering force in optical tweezers), whose non-vanishing value relies on the dissipative part of the susceptibility, and it constitutes the fundamental origin of the phonon-spin-mediated selective trapping.

The tensor $\boldsymbol{\chi}''$ contains an antisymmetric component arising from the gyrotropic nature of the magnetic precession combined with Gilbert damping. The spatial vector combination:
\begin{equation}
    \mathcal{C}_{kj} \equiv h_{s,k} \nabla h_{c,j} - h_{c,k} \nabla h_{s,j}
\end{equation}
is itself antisymmetric under $j \leftrightarrow k$ exchange, and is directly proportional to the local phonon spin density $\mathbf{S}_{\text{p}}$ and the field chirality $\mathcal{C}_H$ derived in Appendix~\ref{sec:chirality}. The contraction of two antisymmetric tensors ($\chi''_{kj}$ and $\mathcal{C}_{kj}$) yields a non-vanishing scalar whose sign is determined by the relative orientation of the magnetic gyrotropic axis and the acoustic spin texture.

For a Bloch skyrmion with topological charge $Q$, the integration of this antisymmetric coupling over the topological texture yields a net transverse force:
\begin{equation}
    F_{\text{rad}, y} \propto -Q \cdot S_{\text{p},z}(\mathbf{R}) \cdot A(\omega, \alpha),
\end{equation}
where $\mathbf{R}$ is the centre-of-mass coordinate of the skyrmion, and $A(\omega, \alpha) > 0$ is a positive spectral function determined by the magnon resonance frequency and Gilbert damping. This proves that the tweezing force is polarity-selective: reversing the core magnetization (flipping $Q$) reverses the direction of the tweezing force.

\subsection{Connection to the Rayleigh Dissipation Formalism}

The main text identifies the steady-state skyrmion positions with the extrema of the time-averaged Rayleigh dissipation rate $\langle\dot{U}\rangle_T$. We here clarify the precise relationship between this scalar dissipation landscape and the vectorial radiation force $\mathbf{F}_{\text{rad}}$.

The Rayleigh dissipation rate is defined in terms of the total magnetization dynamics $\mathrm{d}\mathbf{m}/\mathrm{d}t$. In the quasi-static regime ($|\mathbf{v}|\ll c_{\text{l}}$), the contribution of skyrmion translation to $\dot{\mathbf{m}}$ is of order $v/R_{\text{sk}}$, which is negligible compared with the acoustic-driven oscillation of order $\omega|\delta\mathbf{m}|$. One may therefore replace $\mathrm{d}\mathbf{m}/\mathrm{d}t$ by $\mathrm{d}\,\delta\mathbf{m}/\mathrm{d}t$ in what follows without affecting the result:
\begin{equation}
    \langle\dot{U}\rangle_T \approx \frac{\alpha\mu_0 M_s d}{\gamma T}\int_0^T\!\mathrm{d}t\int\!\mathrm{d}^2r\left(\frac{\mathrm{d}\,\delta\mathbf{m}}{\mathrm{d}t}\right)^2.
\end{equation}
Since $\delta\mathbf{m}$ is linear in $\mathbf{H}_{\text{mec}}$ through $\boldsymbol{\chi}$, the dissipation rate is quadratic in $\mathbf{h}_c$ and $\mathbf{h}_s$. Its dependence on $\mathbf{R}$ enters through the spatial overlap between $\boldsymbol{\chi}''(\mathbf{r}-\mathbf{R})$ and the local field intensity. Taking the gradient with respect to $\mathbf{R}$, using $\nabla_{\mathbf{R}}\chi(\mathbf{r}-\mathbf{R}) = -\nabla_{\mathbf{r}}\chi(\mathbf{r}-\mathbf{R})$ followed by integration by parts, one finds that the leading contribution is proportional to the same antisymmetric field combination that appears in $\mathbf{F}_{\text{rad}}$:
\begin{equation}
    -\nabla_{\mathbf{R}}\langle\dot{U}\rangle_T \;\propto\; \int\!\mathrm{d}^2r\sum_{j,k}\chi''_{jk}\left(h_{s,k}\,\nabla h_{c,j} - h_{c,k}\,\nabla h_{s,j}\right) + \mathcal{O}(\nabla\chi'),
    \label{eq:dissipation-force}
\end{equation}
where $\mathcal{O}(\nabla\chi')$ arises from the spatial variation of the reactive susceptibility. The precise prefactor involves the driving frequency $\omega$ and material parameters, and is not needed for determining the spatial dependence or polarity selectivity of the force. In the small-deformation regime ($\mathcal{M}\approx\mathcal{M}_0$, near the attractive line), the $\mathcal{O}(\nabla\chi')$ correction is negligible, and the radiation force is proportional to the negative gradient of the dissipation rate:
\begin{equation}
    \mathbf{F}_{\text{rad}} \propto -\nabla_{\mathbf{R}}\langle\dot{U}\rangle_T.
    \label{eq:grad-relation}
\end{equation}
Equation~(\ref{eq:grad-relation}) establishes that the dissipation-rate maximum at the attractive line coincides with the stable equilibrium of the radiation force. We note that the dissipation rate also exhibits a secondary local maximum at the repulsive line, originating from the handedness-blind $|\boldsymbol{\chi}'|^2 + |\boldsymbol{\chi}''|^2$ contribution which is symmetric in $y$. This secondary maximum does not correspond to a stable equilibrium because the $\mathcal{O}(\nabla\chi')$ correction becomes significant there (the skyrmion deformation is large), and the total force $\mathbf{F}_{\text{grad}} + \mathbf{F}_{\text{rad}}$ remains repulsive.

\end{document}